\documentclass[onecolumn,nobibnotes,nofootinbib,superscriptaddress]{revtex4}
\usepackage{amsmath,amssymb}
\usepackage[a4paper,bindingoffset=0.2in,left=0.8in,right=0.8in,top=1in,bottom=1in,footskip=.25in]{geometry}
\usepackage{graphicx,subfigure,epsfig}
\usepackage{bigints}

\newcommand{\be}{\begin{equation}}
\newcommand{\ee}{\end{equation}}
\newcommand{\bea}{\begin{eqnarray}}
\newcommand{\eea}{\end{eqnarray}}
\newcommand{\benn}{\begin{displaymath}}
\newcommand{\eenn}{\end{displaymath}}
\newcommand{\beann}{\begin{eqnarray*}}
\newcommand{\eeann}{\end{eqnarray*}}
\newcommand{\oone}{
\begin{picture}(10,8)
\put(5,5){\circle{8}}
\put(2.9,2.5){{\scriptsize 1}}
\end{picture}
}

\newcommand{\vrr}{\mathbf{r}}

\newcommand{\nn}{\nonumber\\}
\newcommand{\dif}{{\rm d}}

\newcommand{\del}{\partial}

\newcommand{\bx}{x_{\bot}}
\newcommand{\by}{y_{\bot}}
\newcommand{\bu}{z'_{\bot}}
\newcommand{\bz}{z_{\bot}}
\newcommand{\abar}{\bar{\alpha}_s}

\newcommand{\Nc}{N_{\rm c}}

\newcommand{\Nf}{N_{\rm f}}

\begin{document}

\title{High energy asymptotic behavior of the $S$-matrix in the saturation region with the smallest dipole running coupling prescription}
\author{Wenchang Xiang}
\email{wxiangphy@gmail.com}
\affiliation{Guizhou Key Laboratory in Physics and Related Areas, Guizhou University of Finance and Economics, Guiyang 550025, China}
\affiliation{Department of Physics, Guizhou University, Guiyang 550025, China}
\author{Yanbing Cai}
\email{myparticle@163.com}
\affiliation{Guizhou Key Laboratory in Physics and Related Areas, Guizhou University of Finance and Economics, Guiyang 550025, China}
\author{Mengliang Wang}
\email{mengliang.wang@mail.gufe.edu.cn}
\affiliation{Guizhou Key Laboratory in Physics and Related Areas, Guizhou University of Finance and Economics, Guiyang 550025, China}
\author{Daicui Zhou}
\email{dczhou@mail.ccnu.edu.cn}
\affiliation{Key Laboratory of Quark and Lepton Physics (MOE) and Institute of Particle Physics, Central China Normal University, Wuhan 430079, China}

%\pacs{}

\begin{abstract}
We present results from analytic solutions to the running coupling, full next-to-leading order, and collinearly improved next-to-leading order Balitsky-Kovchegov equations in the saturation region with the smallest dipole size QCD running coupling prescription. The analytic results of the $S$-matrix of the latter two equations show that the $\exp(-\mathcal{O}(Y^{3/2}))$ rapidity dependence of the solutions are replaced by $\exp(-\mathcal{O}(Y))$ dependence once the running coupling prescription is switched from parent dipole to the smallest dipole prescription, which indicate that the $S$-matrix has a strong dependence on the choice of running coupling prescription. We compute the numerical solutions of these Balitsky-Kovchegov equations with the smallest and parent dipole running coupling prescriptions, the numerical results confirm the analytic outcomes. The rare fluctuations of the $S$-matrix on top of next-to-leading order corrections are also studied under the smallest dipole running coupling prescription in the center of mass frame. It shows that the rare fluctuations are strongly suppressed and less important in the smallest dipole running coupling prescription case as compared to the parent dipole running coupling prescription case.
\end{abstract}

\maketitle

%-----------------------------------------------------------------------------

\section{Introduction}
\label{sec:intro}
Perturbative QCD predicts that the gluon density in high energy hadronic collisions is rapid growth with decreasing Bjorken $x$ (or increasing energy) due to each emitted gluon itself as a source of further emission, which leads to fill up the available phase space for gluon radiation, and forms a state of saturated gluons called the Color Glass Condensate (CGC)\cite{CGC}. The rapidity evolution of the CGC is described by the Balitsky-JIMWLK\footnote{The JIMWLK is the abbreviation of Jalilian-Marian, Iancu, McLerran, Weigert, Leonidov, Kovner.}\cite{B,JIMWLK1,JIMWLK2,JIMWLK3,JIMWLK4} infinite hierarchy of renormalization group equations for the multi-point correlators of Wilson lines which describe a high energy quark or a gluon traveling through a dense target color field. In the mean field approximation, the hierarchy Balitsky-JIMWLK equations decouple and result in a single non-linear integro-differential equation known as the Balitsky-Kovchegov (BK) equation\cite{B,K}. The BK equation is a closed equation, which significantly simplifies the direct applications to phenomenologies, such as proton structure function in deep inelastic scattering at HERA, and particle production in heavy ion collisions at LHC. However, the BK equation resums only large logarithms $\sim\alpha_s\ln(1/x)$ to all orders with a fixed coupling constant $\alpha_s$, thus it is a leading order (LO) equation. It has been found that although the models (i.e. Iancu-Itakura-Munier model\cite{IIM}) inspired by the LO BK equation can give a successfully qualitative fits to the small-$x$ HERA data\cite{GBW,Marquet07,Levin03,Rezaeian14}, there is some tensions when one uses the LO BK equation to quantitatively compare with the experimental data, since the higher order corrections can be very large\cite{Xiang1,Albacete-Kovchegov,Albacete09,IMMST2,Lappi-Mantysaari,Cepila2}.

Over the past decade, the understanding of the non-linear evolution in QCD beyond leading order accuracy has received important developments, which refers to the first calculations of the next-to-leading order (NLO) corrections to the Balitsky-JIMWLK and BK equations\cite{KW,Bnlo}. The NLO corrections to the BK equation were calculated by the resummation of the $\alpha_s N_f$ contributions to all orders with $N_f$ to be as the number of flavors, which allow one to estimate running coupling corrections to the evolution kernel and result in the running coupling Balitsky-Kovchegov (rcBK) equation. In the language of Feynman diagram, this type of NLO corrections refer to the quark loop contributions. Soon after the rcBK equation was derived\cite{KW,Bnlo}, Balitsky and Chirilli in Ref.\cite{Balitsky-Chirilli} found that the gluon loops and the tree gluon diagrams with quadratic and cubic nonlinearities have also significant contributions to the BK evolution equation. By combining the quark part and gluon part contributions, they obtained a full NLO BK evolution equation. However, the first numerical studies to the full NLO BK equation found that the solution strongly depends on the details of the initial condition, and the scattering amplitude decreases with increasing rapidity and can even turn to negative for small dipoles\cite{Lappi-Mantysaari,Lappi-Mantysaari2}, which indicate that the full NLO BK equation is unstable. It was found that the instability comes from a large double transverse logarithm in the BK evolution kernel. Fortunately, the authors in Ref.\cite{IMMST} devised a novel method to resum those single and double collinear logarithms to all orders and got a stablized evolution equation known as collinearly improved BK equation.

The stablized full NLO BK equation was solved numerically and its numerical solution gives a rather good description to the small $x$ HERA data\cite{IMMST2,Albacete17,Cepila2}. However, this numerical process is very cumbersome to use in practice, due to the intricate programming for solving an integro-differential equation and time-costing for running the program. In order to facilitate the use of the BK equation in direct phenomenological application, a lot of efforts were devoted to establish an analytical dipole scattering amplitude in the literature\cite{Levin-Tuchin,Levin16,Levin14,Xiang,Xiang2,Iancu19}. Among them, we analytically solved the full NLO BK equation in the saturation region\cite{Xiang2}. We find that the $\exp(-\mathcal{O}(Y))$ rapidity dependence of the scattering $S$-matrix in the case of running coupling corrections (including only quark loop contributions) is replaced by $\exp(-\mathcal{O}(Y^{3/2}))$ in the full NLO case (with both quark and gluon loop contributions). In addition, the authors in Ref.\cite{Levin16} used another approach to independently solve the full NLO BK equation in the saturation region, and also got $\exp(-\mathcal{O}(Y^{3/2}))$ rapidity dependence of the $S$-matrix. They established a piecewise dipole scattering amplitude based on the analytic solution. It was found that the piecewise amplitude gives a rather successful fits to the small $x$ HERA data. We would like to note that all the calculations mentioned above are used the size of the parent dipole as the argument of the running coupling constant $\alpha_s$. For several years, as we know that the argument of the coupling constant $\alpha_s$ in the NLO BK equation was interpreted to the size of the parent dipole, especially in the applications to phenomenology\cite{Albacete-Kovchegov,Albacete09,Hekki13}, although the authors in Ref.\cite{Balitsky-Chirilli} has pointed out that the proper interpretation of the argument of the coupling constant is the size of the smallest dipole rather than the size of the parent dipole. Recently, It was found that a very good description of the small $x$ HERA data was obtained by using the collinearly improved NLO BK equation with the smallest dipole running coupling (SDRC) prescription among several different prescriptions of QCD running coupling. The significance of the argument of the coupling constant $\alpha_s$ was aroused\cite{IMMST2,Cepila2}.

In this paper, we shall solve analytically the NLO BK equations with the SDRC prescription in the saturation region. To see the significance of the prescription of the running coupling, we firstly recall the analytic solutions of the rcBK, full NLO BK, and collinearly improved NLO BK equations with the parent dipole running coupling (PDRC) prescription in the saturation region, and we shall use these solutions for the latter comparisons. Secondly, we analytically solve the rcBK, full NLO BK, and collinearly improved NLO BK equations again with emphasizing on the SDRC prescription, and we compare the solutions resulting from two different running coupling prescriptions to see how big difference is.

Interestingly, we get that the solutions of the rcBK equation are exactly same under the PDRC and SDRC two different prescriptions, which means that its solution is independent on the choice of the running coupling prescription.
We find that the $\exp(-\mathcal{O}(Y^{3/2}))$ rapidity dependence of the $S$-matrix (solution of the full NLO BK equation) obtained by our previous studies with the PDRC prescription in Ref.\cite{Xiang2} is replaced by $\exp(-\mathcal{O}(Y))$ rapidity dependence once the running coupling prescription is switched to the SDRC prescription, which indicate that the SDRC prescription suppresses the evolution of the dipole scattering amplitude and renders the rapidity dependence of the exponent of the $S$-matrix keeping a linear dependence as the one in rcBK case, see Fig.\ref{smatrix} for a diagrammatic depiction. With the SDRC prescription, we now get that all the solutions of the NLO BK (rcBK, full NLO BK and collinearly improved NLO BK) equations have the same rapidity dependence $\exp(-\mathcal{O}(Y))$.

In order to test these analytic outcomes, we numerically solve the above mentioned NLO BK equations with the focusing on physics in the saturation region. The numerical results support the analytic findings. In addition, we also study the rare fluctuations of the $S$-matrix on top of the full NLO corrections with the SDRC prescription. We find that the rare fluctuation effects take a negligible contribution to the suppression of the evolution of the dipole amplitude, which is unlike the one obtained with the PDRC prescription in our previous publication\cite{Xiang3} where a factor of $\sqrt{2}$ suppression of the exponential factor of the $S$-matrix is occurred when the rare fluctuation effects are included.

%-----------------------------------------------------------
\section{Solutions to the evolution equations with parent dipole running coupling prescription}
To introduce notations and explain the kinematics, we review the rc, full NLO and collinearly improved NLO BK equations and their solutions in the case of PDRC prescription. These solutions will be used for the latter comparisons with the results obtained under the SDRC prescription in the next section.

\subsection{Leading order Balitsky-Kovchegov equation and its solution}

We consider the high energy scattering between a dilute projectile, consisting of a quark-antiquark color dipole with quark leg at transverse coordinate $x_{\bot}$ and an antiquark leg at transverse coordinate $y_{\bot}$, and a dense target which may be another dipole, a hadron or a nucleus. The evolution of the $S$-matrix in the fixed coupling case is described by the BK equation, which we write as\cite{B,K}
\be
\frac{\partial}{\partial Y} S(r, Y) =
           \int d^2r_1 K^{LO}(r, r_1,r_2)
         \left [S(r_1, Y)S(r_2, Y) - S(r, Y) \right],
\label{LOBK}
\ee
where the evolution kernel is given by
\be
K^{LO}(r, r_1, r_2)=\frac{\bar{\alpha}_s}{2 \pi} \frac{r^2}{r_1^2r_2^2} ,
\label{eq_LOBK_K}
\ee
with $\bar{\alpha}_s = \alpha_sN_c/\pi$. Here we use the notation ${\bf r}=x_{\bot}-y_{\bot}$ as the transverse size of parent dipole and ${\bf r}_1=x_{\bot}-z_{\bot}$ and ${\bf r}_2=z_{\bot}-y_{\bot}$ as the transverse sizes of the two emitted daughter dipoles, respectively. The BK equation resums only the leading logarithmic $\alpha_s\ln(1/x)$ in the fixed coupling case, therefore it is a leading-order equation. We would like to note that the BK equation is a mean field version of the Balitsky-JIMWLK hierarchy\cite{JIMWLK1,JIMWLK2,JIMWLK3,JIMWLK4} equations, in which higher order correlations are neglected.

Due to the simple structure of the BK equation, one can analytically solve it in the saturation region, in which the parton density is so high that the dipole scattering amplitude approaches unit, $N\sim1$ ($N=1-S$), thus $S\sim 0$. So, we can neglect the quadratic term of the $S$-matrix in the BK equation in the saturation region, the Eq.(\ref{LOBK}) becomes
\be
\frac{\partial}{\partial Y} S(r, Y) \simeq
           -\int_{1/Q_s}^{r} d^2r_1 K^{LO}(r, r_1, r_2)
           S(r, Y),
\label{eq_LOBK_ap}
\ee
where $Q_s$ is the saturation momentum. The $Q_s$ is an intrinsic momentum scale which provides a separation between dense and dilute parton system. The lower bound of the integral in Eq.(\ref{eq_LOBK_ap}) is set to $1/Q_s$ since the saturation condition requires that the transverse dipole size should be larger than the typical transverse dipole size, $r_s\sim 1/Q_s$. We would like to note that although there are few radiated daughter dipoles having transverse size larger than parent dipole, we still set the upper bound of the integral to $r$ because the kernel has a rapid decay when the size of the daughter dipole is larger than the parent dipole, therefore the contribution from the region, $r_1 > r$, is negligible. It is not hard to find that the integral is governed by the region either from the radiated dipole closing to quark leg of the parent dipole, $1/Q_s\ll |{\bf r_1}|\ll |{\bf r}|$ and $|{\bf r_2}|\sim |{\bf r}|$, or the radiated dipole closing to antiquark leg of the parent dipole, $1/Q_s\ll |{\bf r_2}|\ll |{\bf r}|$ and $|{\bf r_1}|\sim |{\bf r}|$. In this study, we choose to work in the region $|{\bf r_2}|\sim |{\bf r}|$, the Eq.(\ref{eq_LOBK_ap}) becomes
\be
\frac{\partial}{\partial Y} S(r, Y) \simeq
           -2\frac{\bar{\alpha}_s}{2\pi}\pi\int_{1/Q_s^2}^{r^2}dr_1^2 \frac{1}{r_1^2}
           S(r, Y).
\label{eq_LOBK_apf}
\ee
Note that the factor $2$ on the right hand side of Eq.(\ref{eq_LOBK_apf}) comes from the symmetry of the two regions mentioned above. Now we can easily obtain the analytic solution of the LO BK equation in the saturation region by integrating over $r_1$ and $Y$\cite{Levin-Tuchin,Mueller},
\be
S(r, Y)=\exp\left[-\frac{c}{2}\bar{\alpha}_s^2(Y-Y_0)^2\right]S(r, Y_0),
\label{Sol_Kovchegov}
\ee
where we have used $Q_s^2(Y)=\exp\left[c\bar{\alpha}_s(Y-Y_0)\right]Q_s^2(Y_0)$ with $Q_s^2(Y_0)r^2=1$. The Eq.(\ref{Sol_Kovchegov}) shows that in the saturation region the $S$-matrix has a quadratic rapidity dependence in its exponent. It has been found that this $S$-matrix cannot precisely describe the experimental data from HERA\cite{Albacete09}, since the evolution speed of the scattering amplitude $N$ is too fast, in other words the $S$-matrix is too small. So, one needs to take into account the NLO corrections which can make the $S$-matrix becoming larger, or slow down the evolution speed of the dipole scattering amplitude.

%--------------------------------------------------------------

\subsection{Running coupling Balitsky-Kovchegov equation and its solution}
\label{sec:rcBKPDP}
The first NLO corrections to the LO BK equation were derived by Refs.\cite{KW,Bnlo}, which consider the quark loop contributions to the BK evolution kernel and resum $\alpha_s N_f$ to all order. We usually call this modifications as running coupling corrections. The evolution equation including the quark loop corrections reads\cite{Albacete-Kovchegov}
\be
\frac{\partial S(r, Y)}{\partial Y} = \mathcal{R}[S] - \mathcal{S}[S],
\label{rcBK_t}
\ee
where we divide the right hand side of Eq.(\ref{rcBK_t}) into two parts, 'running coupling part' and 'subtraction part'. The 'running coupling part',
\be
\mathcal{R}[S] = \int\,d^2 r_1
  \,K^{\mathrm{rc}}(r, r_1, r_2)
  \left[S(r_1, Y)\,S(r_2, Y)-S(r, Y)\right],
\ee
resumes all power of $\alpha_s N_f$ corrections to evolution kernel, which has the same structure as the LO BK equation but with a modified evolution kernel. The modified kernel has Balitsky type\cite{Bnlo}
\be
K^{\mathrm{rcBal}}(r, r_1, r_2) = \frac{N_c\alpha_s(r^2)}{2\pi^2}\left[\frac{r^2}{r_1^2r_2^2} + \frac{1}{r_1^2}\left(\frac{\alpha_s(r_1^2)}{\alpha_s(r_2^2)}-1\right)
+ \frac{1}{r_2^2}\left(\frac{\alpha_s(r_2^2)}{\alpha_s(r_1^2)}-1\right)\right],
\label{Bal_rc}
\ee
and Kovchegov-Weigert type\cite{KW}
\be
  K^{\mathrm{rcKW}}(r, r_1, r_2)=\frac{N_c}{2\pi^2}\left[
    \alpha_s(r_1^2)\frac{1}{r_1^2}-
    2\,\frac{\alpha_s(r_1^2)\,\alpha_s(r_2^2)}{\alpha_s(R^2)}\,\frac{
      {\bf r}_1\cdot {\bf r}_2}{r_1^2\,r_2^2}+
    \alpha_s(r_2^2)\frac{1}{r_2^2} \right],
\label{KW_rc}
\ee
with
\be
R^2(r, r_1, r_2)=r_1\,r_2\left(\frac{r_2}{r_1}\right)^
{\frac{r_1^2+r_2^2}{r_1^2-r_2^2}-2\,\frac{r_1^2\,r_2^2}{
      {\bf r}_1\cdot{\bf r}_2}\frac{1}{r_1^2-r_2^2}},
\label{R}
\ee
which depend on the decomposition scheme of the two parts, see Ref.\cite{Albacete-Kovchegov} for the details of separation schemes. It is easy to see that the couplings in Eqs.(\ref{Bal_rc}) and (\ref{KW_rc}) are a function of the transverse size of a dipole, which do not like the one in leading order case where the coupling is fixed. We choose the running coupling at one loop accuracy in this study
\be
\alpha_s(r^2) = \frac{1}{b\ln\left(\frac{1}{r^2\Lambda^2}\right)},
\label{runningc}
\ee
with $b=(11N_c-2N_f)/12\pi$. We would like to point out although we have found that the Balitsky and Kovchegov-Weigert kernels reduce to the same one,
\be
K^{\mathrm{rc}}(r, r_1, r_2) = \frac{N_c}{2\pi^2}\frac{\alpha_s(r_1^2)}{r_1^2},
\label{rcKernel_PDP}
\ee
in the saturation region in our previous studies\cite{Xiang}, it has been found by numerical studying the rcBK equation that the Balitsky kernel gives a more reasonable solution as suggested in Ref.\cite{Bnlo}. Therefore, we will only use the Balitsky kernel in the all following studies.

The second term on the right hand side of Eq.(\ref{rcBK_t}) is the 'subtraction part', which has form as\cite{Albacete-Kovchegov}
\bea
\mathcal{S}[S] &=& \alpha_{\mu}^2 \, \int d^2 z_{{\bot}1} \, d^2 z_{{\bot}2} \,
  K_{\oone} (x_{\bot}, y_{\bot} ; z_{{\bot}1}, z_{{\bot}2}) \, \left[ S (x_{\bot}- w_{\bot}, Y) \, S (w_{\bot}-y_{\bot}, Y)\right.\nonumber\\
 &&\hspace*{0.01cm}- \left.S(x_{\bot}-z_{{\bot}1}, Y) \, S(z_{{\bot}2}-y_{\bot}, Y)\right]
\label{sub_term}
\eea
where the $\alpha_{\mu}$ and $w_{\bot}$ are the bare coupling and subtraction point, respectively. The subtraction point can be chosen to be the transverse coordinate of the emitted gluon $z_{\bot}$\cite{KW} or the transverse coordinate of either the quark $z_{{\bot}1}$ or the antiquark $z_{{\bot}2}$\cite{Bnlo}. The evolution kernel in Eq.(\ref{sub_term}) is
written as
\be
 K_{\oone} (x_{\bot}, y_{\bot}; z_{{\bot}1}, z_{{\bot}2})= C_F\sum_{m,n=0}^{1}(-1)^{m+n}\mathcal{K}_{\oone} (x_{\bot m}, y_{\bot n}; z_{{\bot}1}, z_{{\bot}2}),
\label{kernel_sub}
\ee
where the $\mathcal{K}_{\oone} (x_{\bot m}, y_{\bot n}; z_{{\bot}1}, z_{{\bot}2})$ is the resummed JIMWLK kernel. When one substitutes Eq.(\ref{kernel_sub}) into Eq.(\ref{sub_term}), it seems that the Eq.(\ref{sub_term}) is so complicated that the evolution equation, Eq.(\ref{rcBK_t}), cannot be solved analytically. Fortunately, in this study we only focus on the physics in the high-energy region, in which the unitarity corrections become important and the $S$-matrix is small, thus the quadratic terms of the $S$-matrix can be neglected. Keeping the linear term of the $S$-matrix only, the Eq.(\ref{rcBK_t}) simplifies to
\be
\frac{\partial S(r,Y)}{\partial Y}\simeq- 2\int_{1/Q_s}^{r}\,d^2 r_1
  \,\frac{\bar{\alpha}_s(r_1^2)}{2\pi r_1^2}S(r, Y),
\label{SKW_sr}
\ee
where the factor $2$ on the right hand side of Eq.(\ref{SKW_sr}) comes from considering of the symmetry of the two integral regions as the case of fixed coupling. Performing the integral over the dipole
size $r_1$ in Eq.(\ref{SKW_sr}), one gets
\be
\frac{\partial S(r,Y)}{\partial Y}\simeq -\frac{N_c}{b\pi}\left[\ln\left(\ln\frac{Q_s^2}{\Lambda^2}\right)-\ln\left(\ln\frac{1}{r^2\Lambda^2}\right)\right]S(r,Y),
\ee
whose solution is\cite{Xiang}
\be
S(r,Y)=\exp\left\{-\frac{N_c}{bc\pi}\ln^2\frac{Q_s^2}{\Lambda^2}\left[\ln\left(\frac{\ln\frac{Q_s^2}{\Lambda^2}}{\ln\frac{1}{r^2\Lambda^2}}\right)-\frac{1}{2}\right]\right\}S(r,Y_0),
\label{slo_rc}
\ee
with the saturation momentum in the NLO case
\be
\ln\frac{Q_s^2}{\Lambda^2} = \sqrt{c(Y-Y_0)} + \mathcal{O}(Y^{1/6}).
\label{SM}
\ee
We would like to note that in the exponent of the $S$-matrix the rapidity is changed from quadratic dependence in the fixed coupling case, Eq.(\ref{Sol_Kovchegov}), to linear dependence in the running coupling case, Eq.(\ref{slo_rc}). This change significantly slows down the evolution speed of the dipole scattering amplitude, which is supported by the HERA data\cite{Albacete09,Albacete10}.

%-----------------------------------------------------------
\subsection{Full next-to-leading order Balitsky-Kovchegov equation and its solution}
\label{subsec_NLL BK}
 As we know that the rcBK equation only considers part of NLO corrections from quark loop contributions. It has been shown in Ref.\cite{Balitsky-Chirilli} that
 a comprehensive corrections should include both the contributions from the quark and gluon loops as well as from the tree gluon diagrams with quadratic and cubic nonlinearities.
 Including all the above mentioned NLO corrections, one obtains the full NLO BK evolution equation as\cite{Balitsky-Chirilli}
 \bea
 \frac{\partial S(r, Y)}{\partial Y} & = & \frac{\bar{\alpha}_s}{2\pi}\int d^2r_1K_1[S(r_1, Y)S(r_2, Y) - S(r, Y)] + \frac{\bar{\alpha}_s^2}{8\pi^2}\int d^2r_1d^2r_2'K_2[S(r_1, Y)S(r_3, Y)S(r_2', Y)\nonumber\\
 &&- S(r_1,Y)S(r_2,Y)] + \frac{\bar{\alpha}_s^2N_f}{8\pi^2N_c}\int d^2r_1d^2r_2' K_3[S(r_1', Y)S(r_2, Y) - S(r_1, Y)S(r_2, Y)],
 \label{fnlobk}
 \eea
 where the kernels are
 \bea
 K_1 &=& \frac{r^2}{r_1^2 r_2^2}\,
 \bigg[ 1 + \abar
 \bigg(b\, \ln r^2\mu^2
 - b\,\frac{r_1^2 - r_2^2}{r^2}\ln \frac{r_1^2}{r_2^2}
 +\frac{67}{36} - \frac{\pi^2}{12} - \frac{5 \Nf}{18 \Nc}-
 \frac{1}{2}\ln \frac{r_1^2}{r^2} \ln \frac{r_2^2}{r^2}\bigg)
 \bigg], \label{kernel}\\
 K_2 &=& -\frac{2}{r_3^4} + \bigg[\frac{r_1^2r_2'^2 + r_1'^2r_2^2 - 4r^2r_3^2}{r_3^4(r_1^2r_2'^2-r_1'^2r_2^2)} + \frac{r^4}{r_1^2r_2'^2(r_1^2r_2'^2 - r_1'^2r_2^2)} +
 \frac{r^2}{r_1^2r_2'^2r_3^2}\bigg]\ln\frac{r_1^2r_2'^2}{r_1'^2r_2^2},\\
 K_3 &=& \frac{2}{r_3^4} - \frac{r_1'^2r_2^2 + r_2'^2r_1^2 - r^2r_3^2}{r_3^4(r_1^2r_2'^2 - r_1'^2r_2^2)}\ln\frac{r_1^2r_2'^2}{r_1'^2r_2^2}.
 \eea
We use the notation $\vrr = \bx -\by$, $\mathbf{r_1} = \bx -\bz$, $\mathbf{r_2} = \by -\bz$, $\mathbf{r_1'} = \bx -\bu$, $\mathbf{r_2'} = \by -\bu$, and $\mathbf{r_3} = \bz -\bu$, which are the transverse
size of dipoles. In Eq.(\ref{kernel}), the $b$ is the first coefficient of the $\beta$ function, and $\mu$ is the renormalization scale. To simplify the calculation, we firstly absorb the term involving the renormalization scale $\mu$ into the running coupling $\alpha_s$, then the terms involving $b$ are absorbed into $\alpha_s$ by using the Balitsky running coupling scheme which was developed in Ref.\cite{Bnlo}. The kernel $K_1$ can be rewritten as
\bea
K_1 &=& \frac{r^2}{r_1^2r_2^2} + \frac{1}{r_1^2}\bigg(\frac{\alpha_s(r_1^2)}{\alpha_s(r_2^2)}-1\bigg)+\frac{1}{r_2^2}\bigg(\frac{\alpha_s(r_2^2)}{\alpha_s(r_1^2)}-1\bigg)\nn
&& +\frac{\abar(r^2)r^2}{r_1^2r_2^2}\bigg(\frac{67}{36} - \frac{\pi^2}{12} - \frac{5N_f}{18N_c} - \frac{1}{2}\ln\frac{r_1^2}{r^2}\ln\frac{r_2^2}{r^2}\bigg).
\eea
Note that the Balitsky running coupling prescription resums all $\alpha_s\beta$ contributions, especially the term $\sim b\ln\frac{r_1^2}{r_2^2}$.

\begin{figure}[h!]
\setlength{\unitlength}{1.5cm}
\begin{center}
\epsfig{file=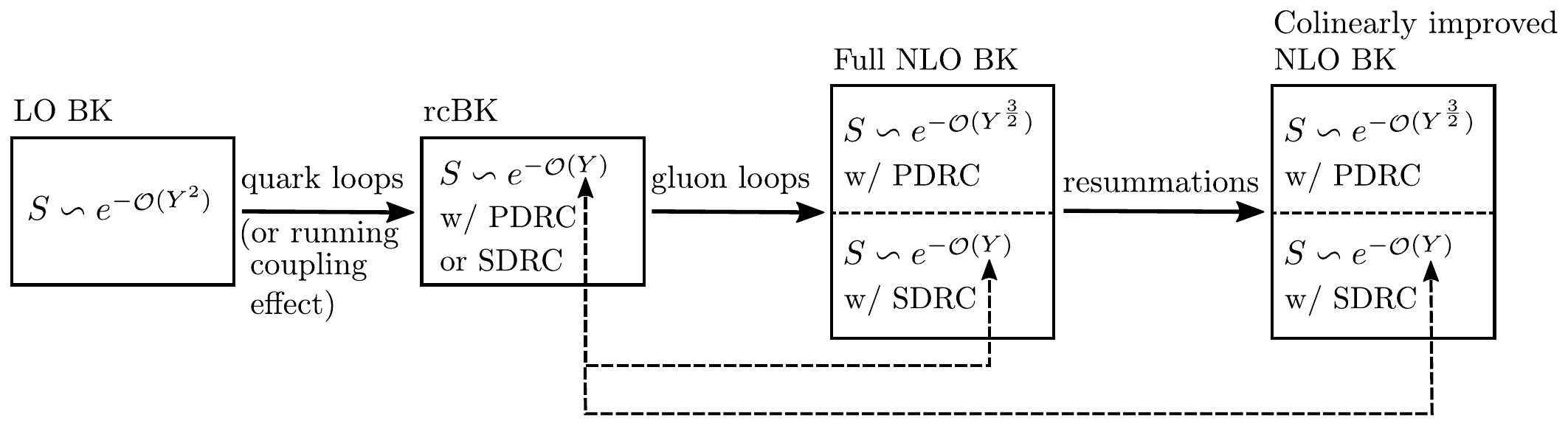, scale=0.7}
\end{center}
\caption{The $S$-matrix from solving the LO, rc, full NLO and collinearly improved NLO BK equations with the PDRC and SDRC prescriptions in the saturation region.}
\label{smatrix}
\end{figure}

We focus on dipole scattering in the saturation region in which the unitarity corrections are very important or $S$ approaches to zero. Thus, one can neglect the non-linear terms in Eq.(\ref{fnlobk}). The full NLO BK evolution equation simplifies to
\be
 \frac{\partial S(r, Y)}{\partial Y} =
 - \frac{\abar(r^2)}{2\pi}\int \dif^2r_1 K_1 S(r, Y).
\label{fnlobk_s}
\ee
Let's turn to analytically solve Eq.(\ref{fnlobk_s}) in either $1/Q_s\ll|\mathbf{r_1}|\ll |\mathbf{r}|,~|\mathbf{r_2}|\sim|\mathbf{r}|$ or $1/Q_s\ll|\mathbf{r_2}|\ll |\mathbf{r}|,~|\mathbf{r_1}|\sim|\mathbf{r}|$ region as mentioned in the LO case. If we choose to work in the first regime, the NLO kernel $K_1$ reduces to
\be
K_1 = \frac{1}{r_1^2}\frac{\alpha_s(r_1^2)}{\alpha_s(r^2)} + \frac{\abar(r^2)}{r_1^2}\left(\frac{67}{36} - \frac{\pi^2}{12} - \frac{5N_f}{18N_c}\right)
\label{kfnlos}
\ee
Substituting the simplified kernel into Eq.(\ref{fnlobk_s}), the evolution equation becomes
\be
 \frac{\partial S(r, Y)}{\partial Y} =
 - 2\frac{1}{2\pi}\int_{1/Q_s}^{r} \dif^2 r_1 \left[\frac{\abar(r_1^2)}{r_1^2} +
\frac{\abar^2(r^2)}{r_1^2}\left(\frac{67}{36} - \frac{\pi^2}{12} - \frac{5N_f}{18N_c}\right)\right] S(r, Y),
 \label{nll_eq}
\ee
and has a solution as following\cite{Xiang2}
\bea
S(r, Y) &=& \exp\Bigg\{-\frac{2BN_c^2}{3b^2c\pi^2}\ln^3\frac{Q_s^2}{\Lambda^2} - \frac{N_c}{bc\pi}\bigg[\ln\bigg(\frac{\ln\frac{1}{r^2\Lambda^2}}{\ln\frac{Q_s^2}{\Lambda^2}}\bigg)-\frac{1}{2}\bigg]\ln^2\frac{Q_s^2}{\Lambda^2}\nn
&& \hspace*{1.0cm} + \frac{B N_c^2\ln\frac{1}{r^2\Lambda^2}}{b^2c\pi^2}\ln^2\frac{Q_s^2}{\Lambda^2}\Bigg\}S(r, Y_0),
\label{nllsol}
\eea
where the NLO saturation momentum has the same definition as Eq.(\ref{SM}) and $B=(67/36-\pi^2/12-5N_f/18N_c)/\ln^2(1/r^2\Lambda^2)$. Let's look at the solutions in LO Eq.(\ref{Sol_Kovchegov}), running coupling Eq.(\ref{slo_rc}), and full NLO Eq.(\ref{nllsol}), and compare the variation of these solutions. We can see that the NLO corrections slow down the evolution speed of the dipole scattering amplitude, the running coupling effect (quark loop contribution) makes the exponent of the $S$-matrix changing from quadratic rapidity dependence to linear rapidity dependence. However, the full NLO effects, which include quark loop and gluon loop contributions, force the linear rapidity dependence back to the rapidity raised to the power of $3/2$ dependence due to gluon loops binging part of compensation to
offset the decrease, see Fig.\ref{smatrix} for a diagrammatic depiction.

%

%-----------------------------------------------------------
\subsection{Collinearly improved next-to-leading order Balitsky-Kovchegov equation and its solution}

We would like to point out that although we get a analytic solution of the full NLO BK equation in the saturation region, it has been found that the numerical solutions of the full NLO BK equation in Ref.\cite{Lappi-Mantysaari} become unstable for many values of initial conditions. The dipole amplitude can decrease with growing energy and can switch to a negative value, which is in disagreement with the theoretical expectations. It has been shown that the main source for such an instability comes from a large double-logarithmic contribution\cite{IMMST}. To solve this unstable problem, one needs to resum double transverse logarithms to all orders under double logarithmic approximation (DLA). There is also a large single transverse logarithms (STL) which appear in the NLO corrections to the BK equation. Such single logarithms must be kept under control to ensure a good convergence of the $\alpha_s$ expansion. The single and double logarithmic resummations were done in Refs.\cite{IMMST,IMMST2}, they gave an collinearly improved evolution equation as
\be
\frac{\partial S(r, Y)}{\partial Y} \simeq \frac{\bar{\alpha}_s}{2\pi}\int d^2r_1K_1^{\mathrm{CI}}[S(r_1, Y)S(r_2, Y) - S(r, Y)],
\label{fnlobk_rsum}
\ee
where the collinearly-improved kernel is\cite{Lappi-Mantysaari2}
\bea
K_1^{\mathrm{CI}} &=& K^{\mathrm{DLA}}K^{\mathrm{STL}}\bigg[\frac{r^2}{r_1^2r_2^2} + \frac{1}{r_1^2}\bigg(\frac{\alpha_s(r_1^2)}{\alpha_s(r_2^2)}-1\bigg)+\frac{1}{r_2^2}\bigg(\frac{\alpha_s(r_2^2)}{\alpha_s(r_1^2)}-1\bigg)\bigg]\nn
&& - \frac{r^2}{r_1^2r_2^2}\bigg(-\abar A_1\bigg|\ln\frac{r^2}{\mathrm{min}\{r_1^2, r_2^2\}}\bigg|\bigg)\nn
&& +\frac{\abar(r^2)r^2}{r_1^2r_2^2}\bigg(\frac{67}{36} - \frac{\pi^2}{12} - \frac{5N_f}{18N_c}\bigg),
\label{ci_kernel}
\eea
with the DLA kernel
\be
K^{\mathrm{DLA}} = \frac{J_1\bigg(2\sqrt{\abar \rho^2}\bigg)}{\sqrt{\abar\rho^2}}\simeq 1 - \frac{\abar\rho^2}{2} + \mathcal{O}(\abar^2),
\ee
and the STL kernel
\be
K^{\mathrm{STL}} = \exp\bigg\{-\abar A_1\bigg|\ln\frac{r^2}{\mathrm{min}\{r_1^2, r_2^2\}}\bigg|\bigg\}.
\ee
The constant $A_1$ in above equations comes from the DGLAP anomalous dimension, the $J_1$ is the Bessel function with $\rho=\sqrt{\ln r_1^2/r^2\ln r_2^2/r^2}$. Note that when $\ln r_1^2/r^2\ln r_2^2/r^2 < 0$, then an absolute value is used and the Bessel function $J_1$ turns into $I_1$\cite{IMMST2}. Here, we would like to note several points, (i) we focus on the physics in saturation region in this paper, thus we neglect the uninteresting terms in Eq.(\ref{fnlobk_rsum}) as compared to Eq.(\ref{fnlobk}); (ii) the double logarithmic resummation is included in the DLA kernel, which is to remove the double logarithmic term from the last line in Eq.(\ref{ci_kernel}) in order to avoid double counting; (iii) the second (subtraction) term in Eq.(\ref{ci_kernel}) is to subtract the $\alpha_s^2$ part of the single transverse logarithm included in $K_2$ to avoid double counting, see Ref.\cite{Lappi-Mantysaari2} for more detailed discussions about the subtraction term.

Now let's turn to analytically solve Eq.(\ref{fnlobk_rsum}) in the saturation region in which the quadratic term of $S$-matrix can be neglected due to very small $S$. As it was done in previous subsections, we choose to work in the region, $1/Q_s\ll|\mathbf{r_1}|\ll |\mathbf{r}|,~|\mathbf{r_2}|\sim|\mathbf{r}|$. In this regime, the $\rho$ is equal to zero leading to the DLA kernel $K^{\mathrm{DLA}}\simeq 1$, which corroborates the statement that the double logarithm only important in phase-space where the scattering is weak\cite{IMMST}. So, we neglect the quadratic term of $S$-matrix and double logarithmic corrections, then the Eq.(\ref{fnlobk_rsum}) becomes
\be
\frac{\partial S(r, Y)}{\partial Y} \simeq -\frac{\bar{\alpha}_s}{2\pi}\int d^2r_1K_1^{\mathrm{CI}}S(r, Y),
\label{fnlobk_rsum_f}
\ee
with a simplified kernel
\bea
K_1^{\mathrm{CI}} &=& K^{\mathrm{STL}}\bigg[\frac{r^2}{r_1^2r_2^2} + \frac{1}{r_1^2}\bigg(\frac{\alpha_s(r_1^2)}{\alpha_s(r_2^2)}-1\bigg)+\frac{1}{r_2^2}\bigg(\frac{\alpha_s(r_2^2)}{\alpha_s(r_1^2)}-1\bigg)\bigg]\nn
&& - \frac{r^2}{r_1^2r_2^2}\bigg(-\abar A_1\ln\frac{r^2}{r_1^2}\bigg)\nn
&& +\frac{\abar(r^2)r^2}{r_1^2r_2^2}\bigg(\frac{67}{36} - \frac{\pi^2}{12} - \frac{5N_f}{18N_c}\bigg),
\label{ci_kernel_f}
\eea
whose solution is
\bea
S(r, Y) &=& \exp\bigg\{-\frac{2BN_c^2}{3b^2c\pi^2}\ln^3\frac{Q_s^2}{\Lambda^2}-\bigg[\frac{N_c}{bc\pi}\bigg(1+\frac{N_c A_1}{b\pi}\bigg)\bigg(\ln\bigg(\frac{\ln\frac{Q_s^2}{\Lambda^2}}{\ln\frac{1}{r^2\Lambda^2}}\bigg)-\frac{1}{2}\bigg)\nn
&&\hspace*{1.0cm}-\frac{N_c^2}{b^2c\pi^2}(A_1+B\ln\frac{1}{r^2\Lambda^2})\bigg]\ln^2\frac{Q_s^2}{\Lambda^2} - \frac{2N_c^2A_1\ln\frac{1}{r^2\Lambda^2}}{b^2c\pi^2}\ln\frac{Q_s^2}{\Lambda^2}\bigg\}S(r, Y_0).
\label{sol_nll_pdp}
\eea
If one compares the two solutions Eq.(\ref{nllsol}) and Eq.(\ref{sol_nll_pdp}), one can find that the dominant terms in the exponent of the $S$-matrix are the same, which indicate that in the saturation region the resummations of double and single transverse logarithms is negligible under the parent dipole running coupling prescription. However, in the next section we shall show that the situation has a dramatic change once the parent dipole running coupling prescription is replaced by the smallest dipole prescription.

%------------------------------------------------------------------------------
\section{Solutions to the evolution equations with the smallest dipole running coupling prescription}
\label{sec:solSDP}
In the last section, we discuss the BK evolution equations with the PDRC prescription. However, the recent studies in Refs.\cite{IMMST2,Cepila1,Cepila2} found that the evolution equation with the SDRC prescription has been advocated to be the correct QCD running coupling prescription, since it is favored by the HERA data at a phenomenological level. It was also pointed out in Ref.\cite{Balitsky-Chirilli} that the proper interpretation of the argument of the QCD coupling is running according to the size of the smallest dipole for the BK equation at NLO. In this section, we start with our discussion on the rcBK evolution equation, since the coupling of the LO BK evolution equation is fixed, thus the argument of the coupling does not affect the solution of the LO BK equation.

%-----------------------------------------------------------
\subsection{Balitsky-Kovchegov equation and its solution in the running coupling case}
\label{subsection_rcSDP}
 As it is shown in Section \ref{sec:rcBKPDP} that the rcBK equation can be written as
 \be
 \frac{\del S(r, Y)}{\del Y} = \int d^2r_1K^{\mathrm{rc}}(r,r_1,r_2)\bigg[S(r_1, Y)S(r_2, Y) - S(r, Y)\bigg],
 \label{rcBK_SDP}
 \ee
with the Balitsky kernel
\be
K^{\mathrm{rc}}(r, r_1, r_2) = \frac{N_c\alpha_s(r^2)}{2\pi^2}\left[\frac{r^2}{r_1^2r_2^2} + \frac{1}{r_1^2}\left(\frac{\alpha_s(r_1^2)}{\alpha_s(r_2^2)}-1\right)
+ \frac{1}{r_2^2}\left(\frac{\alpha_s(r_2^2)}{\alpha_s(r_1^2)}-1\right)\right].
\label{Bal_rcSDP}
\ee
Here, in order to clearly see the effect of the coupling argument we keep only the running coupling part of the evolution equation and neglect the subtraction terms in Eq.(\ref{rcBK_SDP}). Indeed, the subtraction terms are dropped eventually, since they are proportion to the square of the $S$-matrix, while the $S$-matrix is small in this paper's interesting region (saturation regime). We choose to work in the region, $1/Q_s<<|r_1|<<|r|, |r_2|$, as done in previous section, the Balitsky kernel reduces to\cite{Balitsky-Chirilli,Xiang}
\be
K^{\mathrm{rc}}(r, r_1, r_2) = \frac{N_c}{2\pi^2}\frac{\alpha_s(r_1^2)}{r_1^2}.
\label{rcKernel_SDP}
\ee
From the above equation, one can see that the argument of the QCD coupling is the smallest dipole size $r_1$. Interestingly, the kernel in Eq.(\ref{rcKernel_SDP}) is exactly same as the one in Eq.(\ref{rcKernel_PDP}), which implies that the rcBK kernel is independent of the choice of the running coupling prescription in the saturation region. So, the solution of the Eq.(\ref{rcBK_SDP}) is
\be
S(r,Y)=\exp\left\{-\frac{N_c}{bc\pi}\ln^2\frac{Q_s^2}{\Lambda^2}\left[\ln\left(\frac{\ln\frac{Q_s^2}{\Lambda^2}}{\ln\frac{1}{r^2\Lambda^2}}\right)-\frac{1}{2}\right]\right\}S(r, Y_0),
\label{slo_rcSDP}
\ee
which is exactly the same as Eq.(\ref{slo_rc}).

%--------------------------------------------------------------
\subsection{Balitsky-Kovchegov equation and its solution in the full next-to-leading order case}
\label{Sec:bkinfNLOSDP}
In the full NLO case, the BK evolution equation is given by Eq.(\ref{fnlobk}). Here, we focus on the physics in the saturation region in which the $S$-matrix is small, therefore we neglect the quadratic terms of the $S$-matrix, the full NLO evolution equation reduces to,
\be
 \frac{\partial S(r, Y)}{\partial Y} \simeq
 - 2\frac{1}{2\pi}\int_{1/Q_s}^{r} \dif^2 r_1 \left[\frac{\abar(r_1^2)}{r_1^2} +
\frac{\abar^2(r_1^2)}{r_1^2}\left(\frac{67}{36} - \frac{\pi^2}{12} - \frac{5N_f}{18N_c}\right)\right] S(r, Y).
 \label{nll_eq_SDP}
\ee
We need to emphasize that the argument of the QCD coupling in the second term on the right hand side of the above equation is the smallest dipole size $r_1$ rather than the parent dipole size as in Eq.(\ref{nll_eq}). One can see that this change shall significantly affect the behavior of the dipole scattering amplitude.

To solve the Eq.(\ref{nll_eq_SDP}), we use the running coupling at one loop accuracy which is given by Eq.(\ref{runningc}), the equation becomes
\be
\frac{\del S(r, Y)}{\del Y} = - \bigintssss_{1/Q_s^2}^{r^2} dr_1^2\left[\frac{N_c}{b\pi r_1^2\ln\big(\frac{1}{r_1^2\Lambda^2}\big)} + \frac{B'N_c^2}{b^2\pi^2r_1^2\ln^2\big(\frac{1}{r_1^2\Lambda^2}\big)}\right]S(r, Y),
\ee
whose solution is
\be
S(r, Y) = \exp\Bigg\{-\frac{\Nc}{bc\pi}\bigg[\ln\bigg(\frac{\ln\frac{Q_s^2}{\Lambda^2}}{\ln\frac{1}{r^2\Lambda^2}}\bigg)+ \frac{B'\Nc}{b\pi\ln\frac{1}{r^2\Lambda^2}}-\frac{1}{2}\bigg]\ln^2\frac{Q_s^2}{\Lambda^2}
+\frac{2B'\Nc^2}{b^2c\pi^2}\ln\frac{Q_s^2}{\Lambda^2}\Bigg\}S(r,Y_0),
\label{sol_fnlo_SDP}
\ee
where the saturation momentum is given by Eq.(\ref{SM}) and $B'=67/36-\pi^2/12-5\Nf/18\Nc$. By comparing Eq.(\ref{sol_fnlo_SDP}) with Eq.(\ref{nllsol}), one can see that in the exponent of $S$-matrix the rapidity raised to power of $3/2$ dependence is replaced by the linear rapidity dependence once the SDRC prescription is used. The numerical studies in the next section shall prove this finding. We would like to point out that the solution of the full NLO BK equation now has a similar linear rapidity dependence as the one obtained by solving to the rcBK equation with the SDRC prescription, although the evolution speed of the scattering amplitude is slightly rebounded due to the coefficient of the first term in the exponent becoming larger than the running coupling one. The rebound is caused by the gluon loop contributions as we discussed in our previous studies\cite{Xiang2}.

%--------------------------------------------------------------
\subsection{Collinearly improved Balitsky-Kovchegov equation and its solution in the full next-to-leading order case}
In the saturation region, the NLO BK equation with resummation, Eq.(\ref{fnlobk_rsum}), can be rewritten as
\be
\frac{\del S(r, Y)}{\del Y} = - \bigintssss_{1/Q_s^2}^{r^2} dr_1^2\Bigg[\frac{\abar(r_1^2)}{r_1^2}\bigg(1-\abar(r_1^2)A_1\ln\frac{r^2}{r_1^2}\bigg)+\frac{\abar^2(r_1^2)}{r_1^2}\bigg(\frac{67}{36}-\frac{\pi^2}{12}-\frac{5\Nf}{18\Nc}\bigg)\Bigg]S(r, Y)
\label{nll_eq_rsum_SDP}
\ee
under the SDRC description. We need to point out that the argument of the QCD coupling in the above equation is the smallest dipole size $r_1$ instead of parent dipole size $r$, which gives rise to a great impact on the evolution speed of the dipole scattering amplitude. Substituting the one loop running coupling, Eq.(\ref{runningc}), into Eq.(\ref{nll_eq_rsum_SDP}), one gets the evolution equation as
\be
\frac{\del S(r, Y)}{\del Y} = - \bigintssss_{1/Q_s^2}^{r^2} dr_1^2\Bigg[\frac{N_c}{b\pi r_1^2\ln\big(\frac{1}{r_1^2\Lambda^2}\big)} - \frac{\Nc^2A_1}{b^2\pi^2r_1^2\ln^2\big(\frac{1}{r_1^2\Lambda^2}\big)} + \frac{B'\Nc^2}{b^2\pi^2r_1^2\ln^2\big(\frac{1}{r_1^2\Lambda^2}\big)}\Bigg]S(r, Y),
\label{nll_eq_rsum_SDPrc}
\ee
whose solution is
\bea
S(r, Y) &=& \exp\Bigg\{-\frac{\Nc}{bc\pi}\bigg[\Big(1-\frac{\Nc A_1}{b\pi}\Big)\ln\bigg(\frac{\ln\frac{Q_s^2}{\Lambda^2}}{\ln\frac{1}{r^2\Lambda^2}}\bigg)+ \frac{B'\Nc}{b\pi\ln\frac{1}{r^2\Lambda^2}}-\frac{1}{2}-\frac{3\Nc A_1}{2b\pi}\bigg]\ln^2\frac{Q_s^2}{\Lambda^2}\nn
&&\hspace*{1.0cm}+\frac{2\Nc^2}{b^2c\pi^2}\Big(B'-A_1\ln\frac{1}{r^2\Lambda^2}\Big)\ln\frac{Q_s^2}{\Lambda^2}\Bigg\}S(r,Y_0).
\label{sol_nll_SDP}
\eea
This result deserves several important comments,
\begin{itemize}
  \item One can see that in the exponent the rapidity raised to power of $3/2$ dependence with the PDRC prescription, Eq.(\ref{sol_nll_pdp}), is now replaced by linear rapidity dependence under the SDRC prescription, which indicates that the argument of the coupling plays an important role on the rapidity dependence of the $S$-matrix.
  \item By comparing the coefficients of the dominant terms between Eq.(\ref{sol_fnlo_SDP}) and Eq.(\ref{sol_nll_SDP}), one can find that in the saturation region the resummation of single transverse logarithms takes a significant effect on the suppression of the evolution of the scattering amplitude, although the resummation of the double logarithmic corrections has a negligible effect on it. This finding is supported by the numerical results performed in the next section.
  \item We can see that under the SDRC prescription all the solutions of the NLO BK equation, Eqs.(\ref{slo_rcSDP}), (\ref{sol_fnlo_SDP}), and (\ref{sol_nll_SDP}), have linear rapidity dependence in the exponent of the $S$-matrix, while under the parent dipole prescription only the solution of the rcBK equation, Eq.(\ref{slo_rc}), has linear rapidity dependence and the other two solutions, Eqs.(\ref{nllsol}), and (\ref{sol_nll_pdp}), have the rapidity raised to power of $3/2$ dependence, see Fig.\ref{smatrix} for a diagrammatic depiction. These results imply that the running coupling corrections (quark loop corrections) is a dominant effect over all the other NLO corrections, like gluon loop corrections, in the suppression of the evolution of the dipole scattering amplitude in the saturation region.
\end{itemize}

To conclude, we can see from the above discussion that the prescription of the QCD running coupling has a strong impact on the rapidity dependence of the $S$-matrix. The SDRC prescription takes a dramatic suppression on the evolution speed of the dipole scattering amplitude, which is favored by HERA data\cite{Cepila1,Cepila2} and satisfies the theoretical expectations\cite{Balitsky-Chirilli}.

%------------------------------------------------------------------------------
\section{Numerical analysis}
In this section, we perform the numerical studies to the NLO BK evolution equations in order to test the analytic solutions obtained in the above sections. The translational invariant approximation is used in our numerical studies, we suppose that the scattering matrix is independent of the impact parameter of the collision, $S =S(|\textbf{r}|, Y)$. The evolution equations are complicated integro-differential equations, but they can numerically straightforward solve on a lattice. The variable $r$ is discretized into 800 points which are equally separated in the logarithmic space between $r_{min}=10^{-8} \mathrm{GeV^{-1}}$ and $r_{max}=50\mathrm{GeV^{-1}}$.    We solve these equations by using the GNU Scientific Library (GSL). To be more specific, the Runge-Kutta method with a step size in rapidity $\Delta Y =0.1$ is used to solve the differential equations, all the integrals are performed using adaptive integration routines, and the cubic spline interpolation method is employed to interpolate the data points.

To solve the evolution equations, we use the McLerran-Venugopalan (MV) model as the initial condition\cite{MV},
\be
N^{\mathrm{MV}}(r, Y=0) = 1 - \exp\bigg[-\frac{r^2Q_s^2(Y)}{4}\ln\Big(\frac{1}{r^2\Lambda^2}+e\Big)\bigg],
\ee
where we set $Q_s(Y)=1$ at $Y=0$, and put $\Lambda=0.2\mathrm{GeV}$. For the running coupling $\alpha_s(r^2)$ in the evolution equations, we use the one-loop running coupling, Eq.(\ref{runningc}), with $\Nf=3$. We freeze the coupling to a fixed value $\alpha_s(r_{\mathrm{fr}})=0.75$ for larger dipole size, $r>r_{\mathrm{fr}}$, to regularize the infrared behavior.

\begin{figure}[h!]
\setlength{\unitlength}{1.5cm}
\begin{center}
\epsfig{file=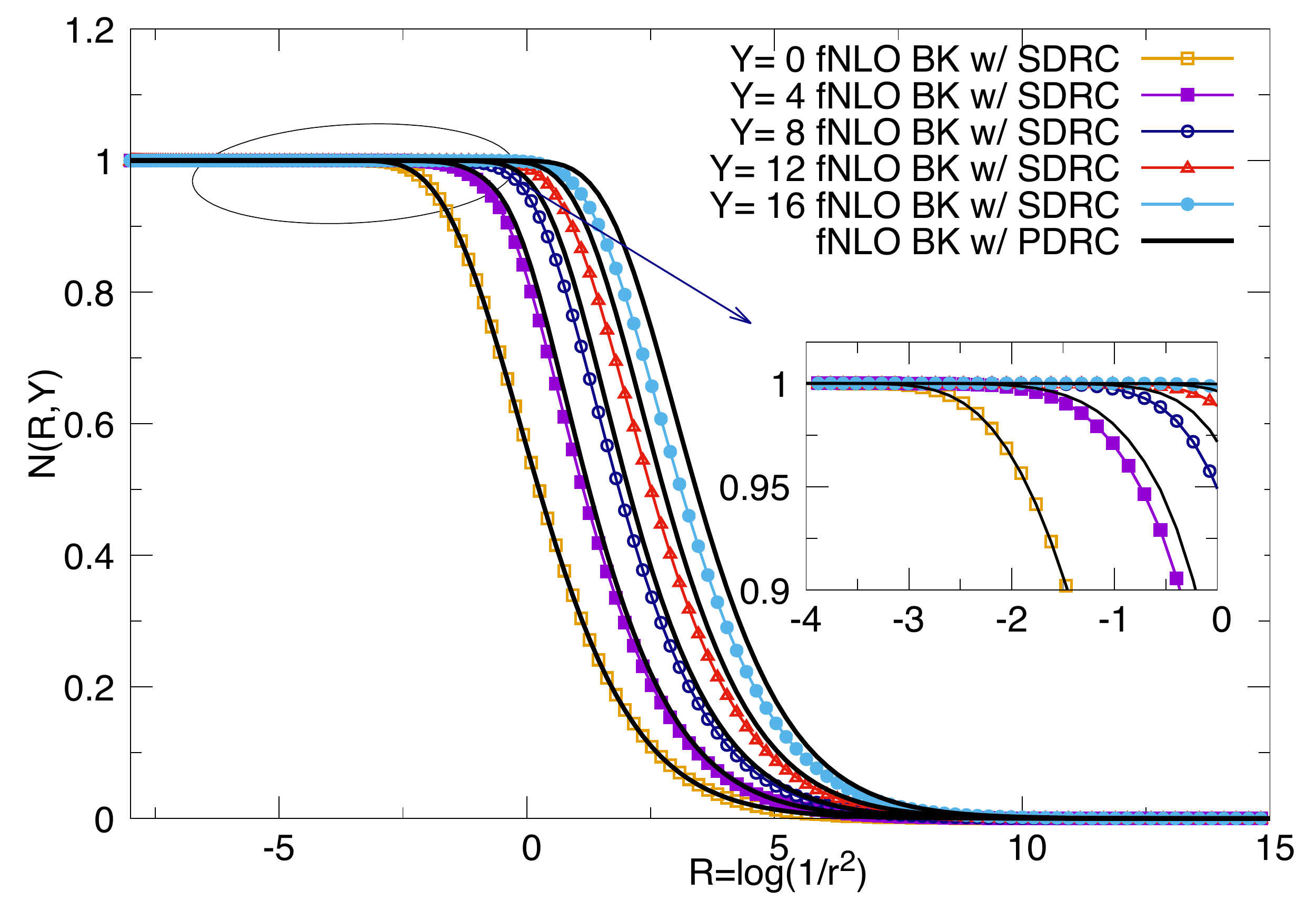, width=8cm,height=7cm}
\epsfig{file=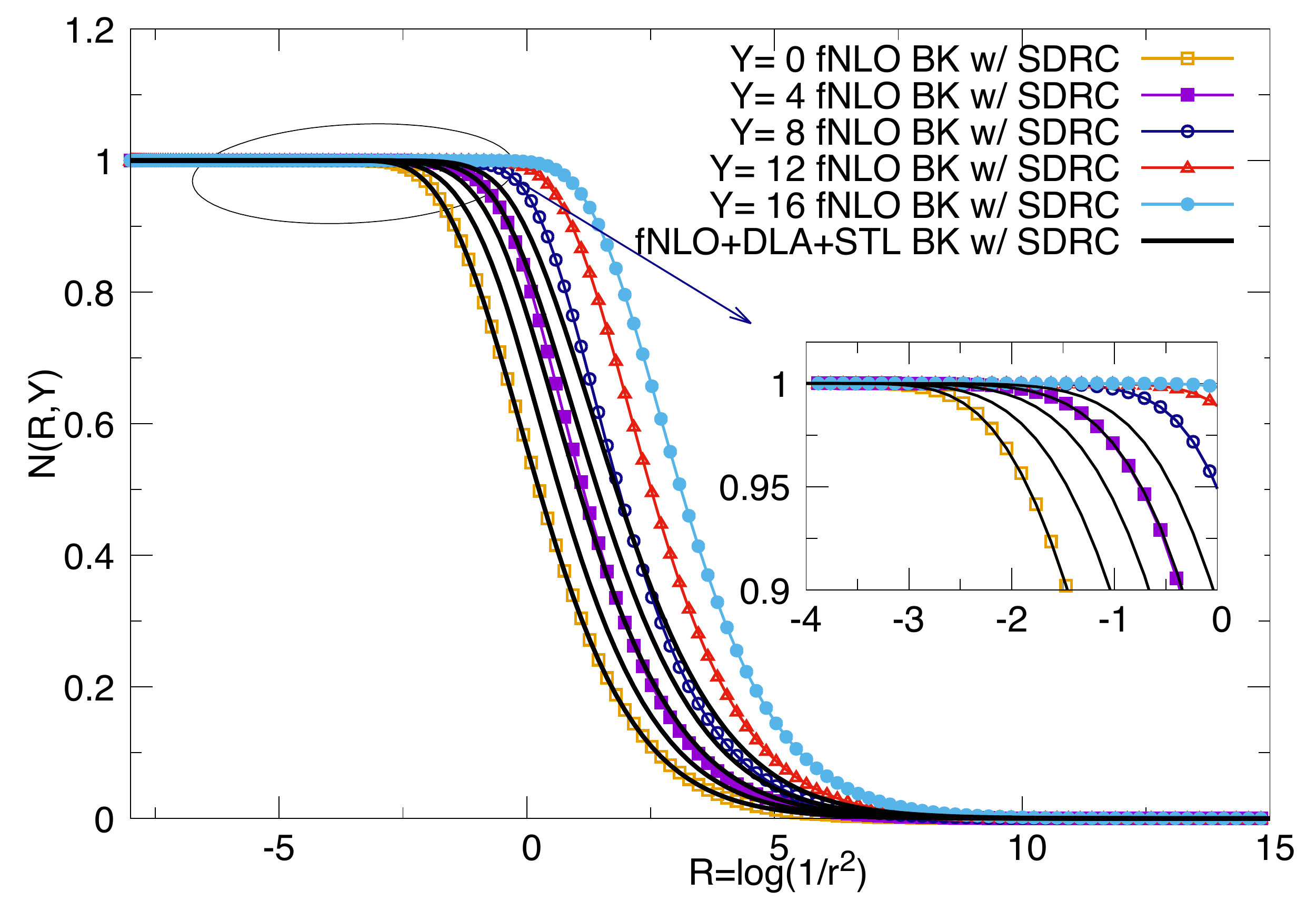, width=8cm,height=7cm}
\end{center}
\caption{Numerical solutions to the full NLO BK equations with the PDRC and the SDRC prescriptions for 5 different rapidities. The left hand panel gives the comparison of the amplitudes between two different running coupling prescriptions. The right hand panel gives the comparison of the amplitudes with and without resummation corrections. The inner diagrams are the zooming in amplitudes in the saturation region.}
\label{figdlanlo}
\end{figure}

The left hand panel of Fig.\ref{figdlanlo} shows the solutions of the full NLO BK equation as a function of the dipole transverse size in the cases of PDRC and SDRC prescriptions for 5 different rapidities. To clearly show the numerical solutions in the saturation region, we plot a zooming in diagram in Fig.\ref{figdlanlo}.
By comparing the full NLO dipole scattering amplitudes with these two different prescriptions, we can see that the amplitudes with the SDRC prescription are smaller than the ones with the PDRC prescription, respectively. This numerical result is in agreement with the analytic finding, Eq.(\ref{sol_fnlo_SDP}) in Section \ref{Sec:bkinfNLOSDP}, in which the rapidity raised to power of $3/2$ dependence in the exponent of the $S$-matrix is replaced by the linear rapidity dependence once the SDRC prescription is used. The right hand panel of Fig.\ref{figdlanlo} gives the comparison between the solutions of full NLO BK equation and collinearly improved NLO BK equation (including a resummation of large single and double transverse logarithms) with the SDRC prescription for 5 different rapidities. We would like to note that the inner zooming in diagram is to clearly show the corresponding solutions in the saturation region. One can see that the evolution of the dipole scattering amplitude is significantly slowed down by the resummation corrections. This outcome supports the analytic results in Section \ref{Sec:bkinfNLOSDP}, where the amplitude, Eq.(\ref{sol_nll_SDP}), is suppressed by the resummation corrections as compared to the one, Eq.(\ref{sol_fnlo_SDP}), without resummations.

\begin{figure}[h!]
\setlength{\unitlength}{1.5cm}
\begin{center}
\epsfig{file=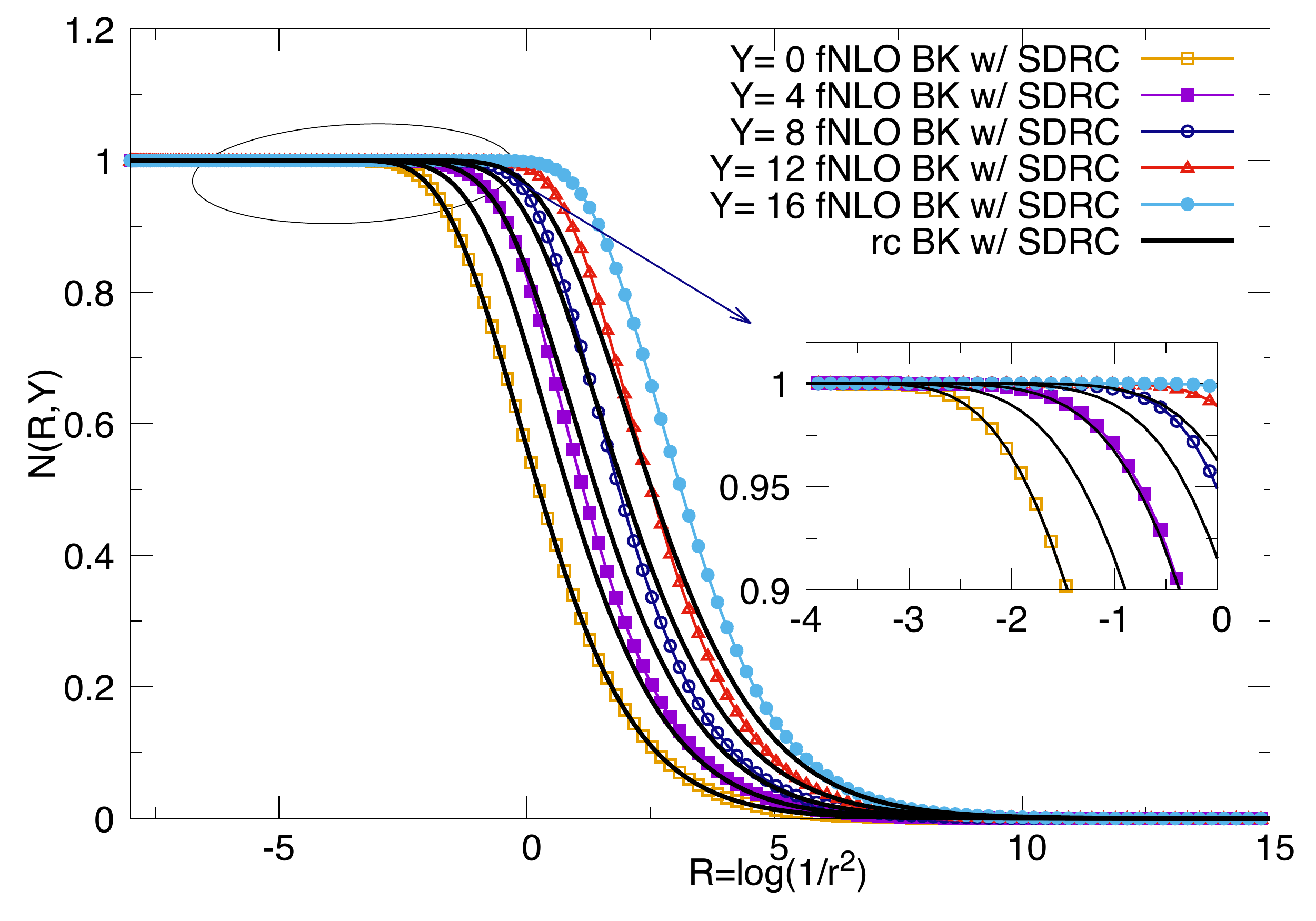, width=8cm,height=7cm}
\epsfig{file=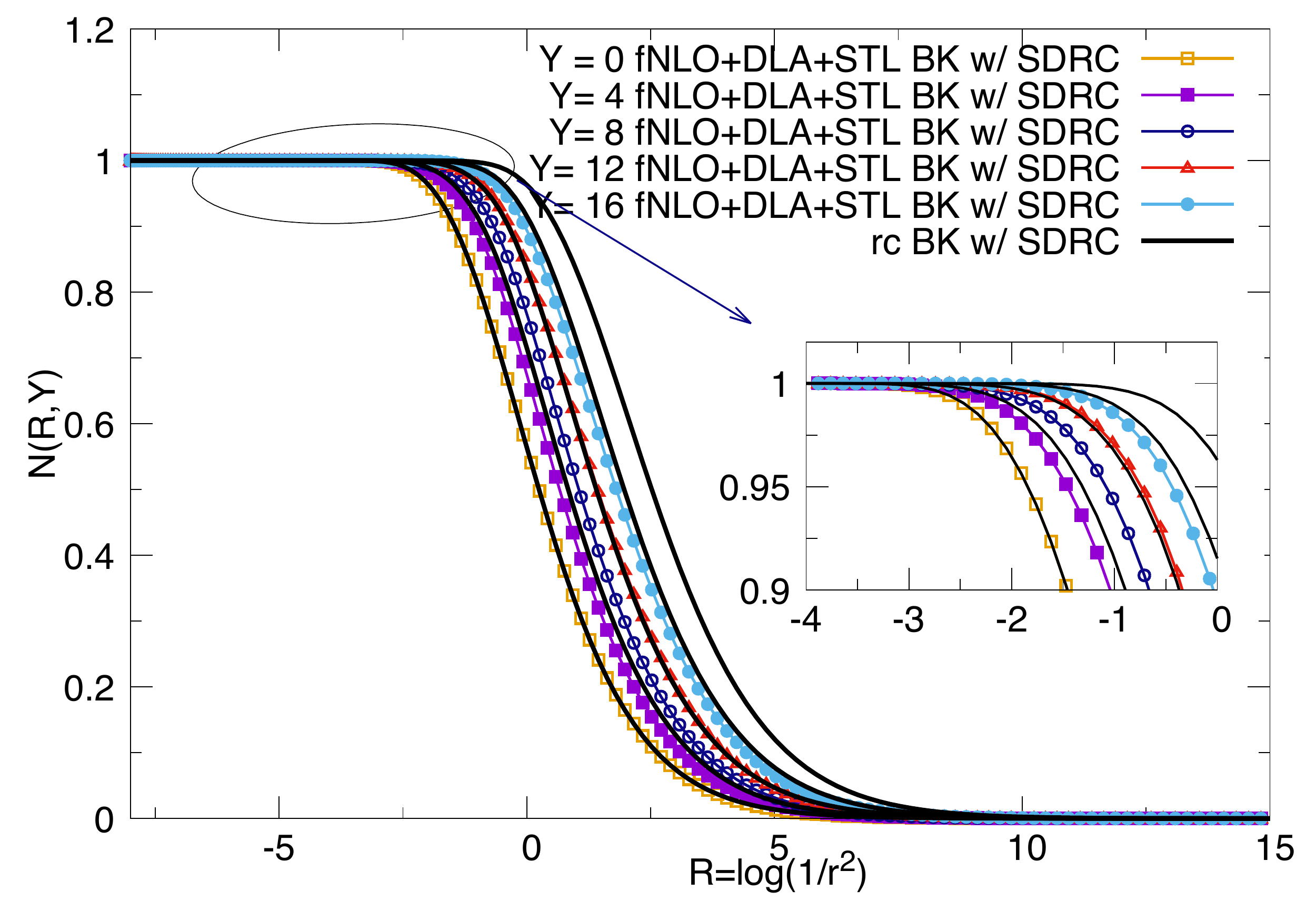, width=8cm,height=7cm}
\end{center}
\caption{Numerical solutions to the rc and full NLO BK equations with the SDRC prescription for 5 different rapidities. The left hand panel gives the comparison of the amplitudes with the running coupling and full NLO corrections. The right hand panel gives the comparisons between the collinearly improved amplitudes and running coupling modified amplitudes. The inner diagrams are the zooming in amplitudes in the saturation region.}
\label{figrcsdp}
\end{figure}

The left hand panel of Fig.\ref{figrcsdp} gives the solutions of the rc and full NLO BK equations as a function of the dipole size with the SDRC prescription for 5 different rapidities. By comparing the corresponding dipole scattering amplitudes between the rc and full NLO cases for each rapidity, we can see that the full NLO scattering amplitude is larger than the running coupling one, which indicates that the gluon loop effect compensates part of decrease made by quark loop effect (running coupling effect). The outcomes is similar as the findings in our previous paper\cite{Xiang2} in which the solution was calculated with the PDRC prescription. One can see that the gluon loop has a rebound effect regardless of which running coupling prescription is used. However, we would like to point out that the compensation to the decrease of the dipole scattering amplitude in the SDRC prescription is less than the parent dipole case. In other words, the SDRC prescription slows down the evolution speed of the dipole scattering amplitude more than the PDRC prescription, which is consistent with the phenomenological desire\cite{IMMST2}. The right hand panel of Fig.\ref{figrcsdp} shows the dipole scattering amplitudes in the rcBK case and the collinearly improved NLO BK (resummations of single and double transverse logarithms) for 5 different rapidities in the case of the SDRC prescription. If one compares the corresponding amplitudes between two cases respectively, one finds that the amplitude is dramatically suppressed by the resummation corrections, especially by the resummation of single logarithm. The resummation of the single logarithm takes effect not only in the non-saturation region but also in the saturation region (see the zooming in diagram on the right hand panel of Fig.\ref{figrcsdp}), it is unlike the resummation of double logarithm which takes a negligible effect in the saturation region\cite{Xiang2}.

%-----------------------------------------------------------
\section{Rare fluctuations of the $S$-matrix with the smallest dipole running coupling prescription}
\label{sec_rare}
As we have studied in Refs.\cite{Xiang,Xiang3} that the rare fluctuations can play an important effect in the suppression of the evolution speed of the dipole scattering amplitude in the saturation region. It was shown that in the case of fixed coupling (LO) the exponential factor of the $S$-matrix without the rare fluctuation effect is twice as large as the result when the rare fluctuations are taken into account\cite{Iancu-Mueller}. We also demonstrated that in the saturation regime the exponential factor of the full NLO $S$-matrix is $\sqrt{2}$ as large as the result which emerges when the rare fluctuation effects are considered. Although it seems that the rare fluctuations are important in the LO and full NLO BK cases, our studies showed that in the rcBK case the rare fluctuations take a tiny effect in the suppression of the evolution speed of the dipole scattering amplitude, thus they can be ignored. We need to point out that all just aforementioned running couplings are used the PDRC prescription. In this study, we want to see if one switches the QCD running coupling prescription of the NLO BK equations to the SDRC prescription, whether the rare fluctuations are still important or not in the suppression of the evolution speed of the dipole amplitude. In this section, we only present the rare fluctuations of the $S$-matrix on top of the full NLO effect in the center of mass (CM) frame, since the relevant results are exactly the same when one works in a general frame, please see the Appendix A.

\begin{figure}[h!]
\setlength{\unitlength}{1.5cm}
\vspace{0.5cm}
\begin{center}
\epsfig{file=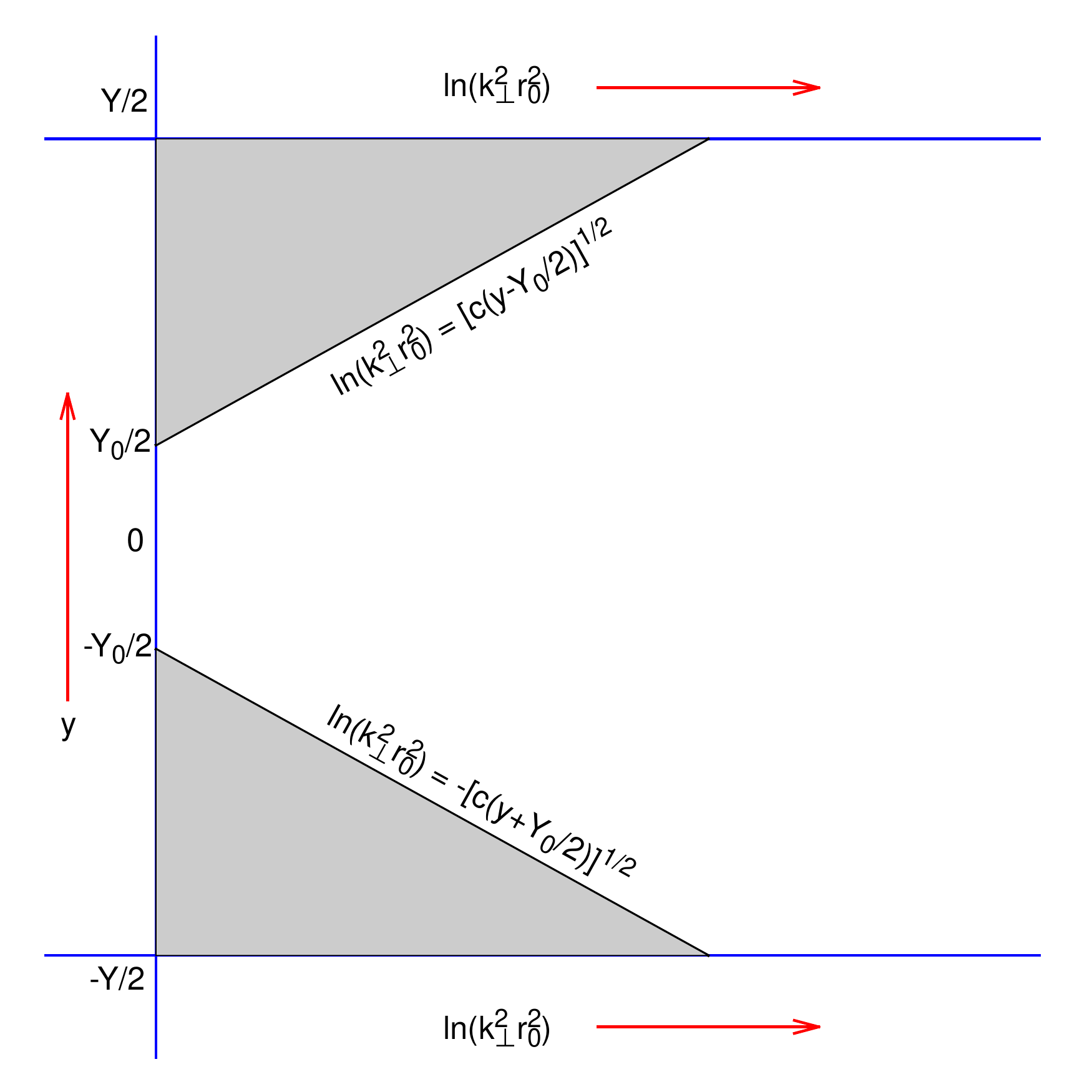, scale=0.5}
\end{center}
\vspace{0.2cm}
\caption{A next-to-leading order rare configuration in the center of mass frame.}
\label{figNLOCMF}
\end{figure}

Following the framework of Ref.\cite{Iancu-Mueller}, we consider a high energy scattering of a left-moving dipole on a right-moving dipole at zero impact parameter in the center of mass frame. We let the right-moving and left-moving dipoles have normal Balitsky-Fadin-Kuraev-Lipatov (BFKL) evolution\cite{BFKL1,BFKL2} in the rapidity intervals $0<y<Y_0/2$ and $-Y_0/2<y<0$, respectively. In order to produce the rare configuration, we need to constrain the rapidity evolution of the system in such a way that the wavefunctions of the right-moving and left-moving dipoles consist of only parent dipole with size $r_0$ in the rapidity intervals $Y_0/2<y<Y/2$ and $-Y/2<y<Y_0/2$, respectively. However, the aforemention constraint of the evolution is a optimal case which cannot be obtained in a real dipole scattering, since one cannot require all the evolutions are absent. What we can do is to only allow the evolutions to produce dipoles with size much smaller than $r_0$ in order to avoid the daughter dipoles evolving into ones with similar size as $r_0$ in intermediate rapidities, see Fig.\ref{figNLOCMF}. We require that the gluon emission from the parent dipoles, as part of the evolution which forms the left and right moving states which scatter on each other, is forbidden if the gluon has $k_{\perp}$ and $y$ lying in the shaded triangles of Fig.\ref{figNLOCMF}. The upper line in Fig.\ref{figNLOCMF} is determined by the requirement that gluons on the right hand side of that line cannot evolve by normal BFKL evolution into shaded triangle,
\be
\ln(k_{\perp}r_0^2) = \sqrt{c\bigg(y-\frac{Y_0}{2}\bigg)},
\ee
and the same requirement is applied to the lower line.

According to the above description, one knows that the probability of rare configuration $\mathbb{S}(r,Y-Y_0)$ satisfies the evolution equation\cite{Iancu-Mueller,Xiang,Xiang3}
\be
  \frac{\partial}{\partial Y}\mathbb{S}(r,Y-Y_0)=-\int d^2r_1 K^{\mathrm{fNLO}}({\bf r},{\bf r}_1,{\bf r}_2)\mathbb{S}(r,Y-Y_0),
\label{fnloBKrare}
\ee
with the kernel
\be
K^{\mathrm{fNLO}} = \frac{\abar(r_1^2)}{2\pi r_1^2} +
\frac{\abar^2(r_1^2)}{2\pi r_1^2}\left(\frac{67}{36} - \frac{\pi^2}{12} - \frac{5N_f}{18N_c}\right),
\ee
whose solution is
\be
\mathbb{S}(r,Y-Y_0) = \exp\Bigg\{-\frac{\Nc}{bc\pi}\bigg[\ln\bigg(\frac{\ln\frac{Q_s^2}{\Lambda^2}}{\ln\frac{1}{r^2\Lambda^2}}\bigg)+ \frac{B'\Nc}{b\pi\ln\frac{1}{r^2\Lambda^2}}-\frac{1}{2}\bigg]\ln^2\frac{Q_s^2}{\Lambda^2}
+\frac{2B'\Nc^2}{b^2c\pi^2}\ln\frac{Q_s^2}{\Lambda^2}\Bigg\},
\label{sol_fnlo_rf}
\ee
with $B'=67/36-\pi^2/12-5\Nf/18\Nc$. Note that Eq.(\ref{sol_fnlo_rf}) is derived by using the SDRC prescription instead of the PDRC prescription, since we focus on studying the rare fluctuation effects of the $S$-matrix in the case of the SDRC prescription in this study.

To obtain the $S$-matrix including the rare fluctuations in the center of mass frame, we need to compute
\be
S(r, Y) = \mathbb{S}\bigg(r,\frac{Y-Y_0}{2}\bigg)\mathbb{S}\bigg(r,\frac{Y-Y_0}{2}\bigg)S(r, Y_0),
\ee
then the $S$-matrix can be calculated as
\bea
S(r, Y) &=& \mathbb{S}\left(r, \frac{Y-Y_{0}}{2}\right)\mathbb{S}\left(r, \frac{Y-Y_{0}}{2}\right)S(r, Y_{0})\nn
&=& \exp\Bigg\{-\frac{2\Nc}{bc\pi}\bigg[\ln\bigg(\frac{\sqrt{\frac{c(Y-Y_0)}{2}}}{\ln\frac{1}{r^2\Lambda^2}}\bigg)+ \frac{B'\Nc}{b\pi\ln\frac{1}{r^2\Lambda^2}}-\frac{1}{2}\bigg]\frac{c(Y-Y_0)}{2}
+\frac{4B'\Nc^2}{b^2c\pi^2}\sqrt{\frac{c(Y-Y_0)}{2}}\Bigg\}\nn
&&\hspace*{0.7cm}\times S(r, Y_0)\nn
&=& \exp\Bigg\{-\frac{\Nc}{bc\pi}\bigg[\ln\bigg(\frac{\frac{\sqrt{2}}{2}\ln\frac{Q_s^2}{\Lambda^2}}{\ln\frac{1}{r^2\Lambda^2}}\bigg)+ \frac{B'\Nc}{b\pi\ln\frac{1}{r^2\Lambda^2}}-\frac{1}{2}\bigg]\ln^2\frac{Q_s^2}{\Lambda^2}
+\frac{2\sqrt{2}B'\Nc^2}{b^2c\pi^2}\ln{\frac{Q_s^2}{\Lambda^2}}\Bigg\}\nn
&&\hspace*{0.7cm}\times S(r, Y_0).
\label{SNLLrareres}
\eea
If one compares Eq.(\ref{SNLLrareres}) with Eq.(\ref{sol_fnlo_SDP}), one can find that the dominant terms in the exponent of the $S$-matrix are almost same, which indicate that the rare fluctuations are not important in the case of the SDRC prescription. Thus, the rare fluctuation effects can be neglected when one works with the SDRC prescription. It is simple to find that the rare fluctuations are also not important in the collinearly improved NLO BK case if one uses the SDRC prescription instead of the PDRC prescription. However, it is necessary to remind that we found a $1/\sqrt{2}$ suppression to the exponential factor of the $S$-matrix when the rare fluctuations are taken into account in our previous studies\cite{Xiang3}, where the PDRC prescription was used.

%-----------------------------------------------------------------------------

\begin{acknowledgments}
WX thanks the Department of Physics of Indiana University Bloomington for the hospitality during the early stages of this work. This work is supported by the National Natural Science Foundation of China under Grant Nos.11765005, 11305040, and 11847152; the Fund of Science and Technology Department of Guizhou Province under Grant Nos.[2018]1023, and [2019]5653; and the Education Department of Guizhou Province under Grant No.KY[2017]004.
\end{acknowledgments}

%-----------------------------------------------------------------------------
%

\appendix
\section{Rare fluctuations of the $S$-matrix in a general frame}

\begin{figure}[h!]
\setlength{\unitlength}{1.5cm}
\vspace{0.5cm}
\begin{center}
\epsfig{file=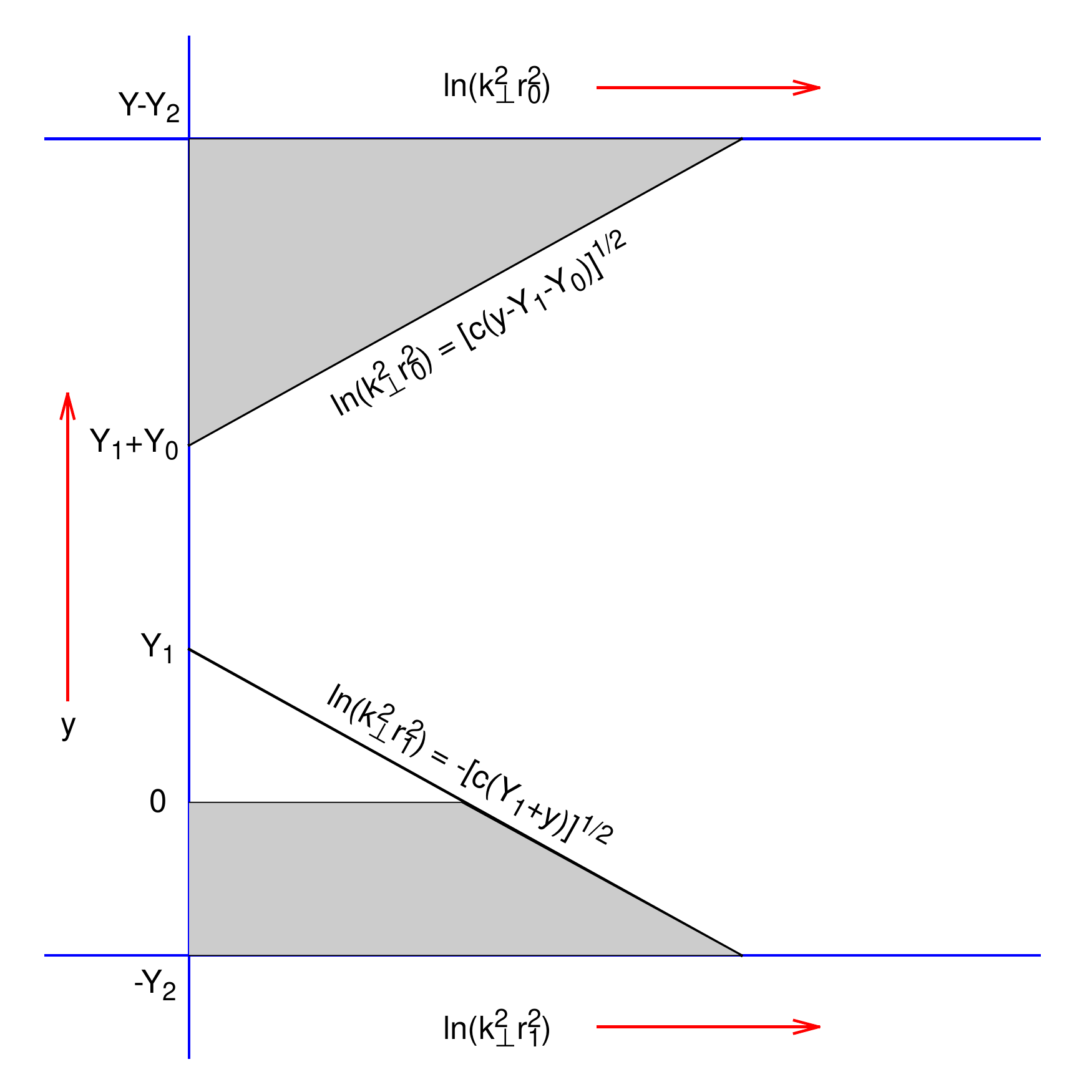, scale=0.5}
\end{center}
\vspace{0.2cm}
\caption{A next-to-leading order rare configuration in a general frame.}
\label{figNLOGF}
\end{figure}

To ensure the relevant results of the $S$-matrix are independent of the frame choice, we study the rare fluctuations of the $S$-matrix on top of the full NLO corrections with the SDRC description in a general frame. We let a right-moving dipole with size $r_0$ and rapidity $Y-Y_2$ scattering off a left-moving dipole with size $r_1$ and rapidity $-Y_2$, see Fig.\ref{figNLOGF} for the scattering picture. For later convenience, we assume $Y_2\leq (Y-Y_0)/2$, where $Y_0$ is the rapidity difference between the two scattering dipoles at the 'time' of the onset of unitarity corrections. Since we suppose that $Y_2\leq (Y-Y_0)/2$, which indicates that the left-moving dipole takes the smaller rapidity. Thus, it is easy to suppress its evolution. We do not allow any additional daughter dipoles created by gluon emission, which means that we have to suppress the emission of those dipoles which, after a normal BFKL evolution during the intermediate rapidities $-Y_2<y<0$, could become of size $1/Q_s$ or larger. Therefore, we require that the gluon emission is forbidden if the gluon has $k_{\perp}$ and $y$ lying in the lower shaded triangle of Fig.\ref{figNLOGF}. For the right-moving dipole, we suppress the evolution in such a way that the gluons laying on the right hand side of the line
\be
\ln(k_{\perp}^2r_0^2) = \sqrt{c(y-Y_1-Y_0)}
\ee
cannot evolve into the upper shaded triangle in Fig.\ref{figNLOGF} through normal BFKL evolution in the rapidity interval $Y_1+Y_0 < y < Y-Y_2$. We let the parent dipole undergo a normal evolution during $Y_1< y <Y_0+Y_1$, where $Y_1$ is determined by maximizing the $S$-matrix later. The unshaded triangle, $0<Y<Y_1$, denotes the saturation region where the right-moving dipole has already evolved into a CGC state.

In terms of the scattering picture depicted above, we can calculate the the rare fluctuations of the $S$-matrix on top of the NLO corrections as\cite{Xiang3}
\be
S(r_0,r_1,Y)=\mathbb{S}_R(r_0,Y-Y_0-Y_1-Y_2)\mathcal{S}(r_0,r_1,Y_0+Y_1)\mathbb{S}_L(r_1,Y_2),
\label{Srareloge}
\ee
where the $\mathcal{S}$ is the $S$-matrix of a elementary dipole scattering off a CGC. The $\mathcal{S}$ can be computed by using the full NLO BK equation with SDRC prescription. After using Eq.(\ref{sol_fnlo_SDP}), one get
\be
\mathcal{S}(r_0,r_1,Y_0+Y_1) = \exp\Bigg\{-\frac{\Nc}{bc\pi}\bigg[\ln\bigg(\frac{\sqrt{cY_1}}{\ln\frac{1}{r^2\Lambda^2}}\bigg)+ \frac{B'\Nc}{b\pi\ln\frac{1}{r^2\Lambda^2}}-\frac{1}{2}\bigg]cY_1
+\frac{2B'\Nc^2}{b^2c\pi^2}\sqrt{cY_1}\Bigg\}S(r,Y_0).
\label{Sdipole_gf}
\ee
Now we turn to compute the suppression factors, $\mathbb{S}_R$, and $\mathbb{S}_L$, which denote no gluon emission from the right-moving and left-moving dipoles, and can be estimated in terms of the area of the upper and lower shaded triangles in Fig.\ref{figNLOGF}\cite{Iancu-Mueller,Xiang3},
\bea
\mathbb{S}_R(r_0,Y-Y_0-Y_1-Y_2) &=& \exp\Bigg\{-\frac{\Nc}{bc\pi}\bigg[\ln\bigg(\frac{\sqrt{c(Y-Y_0-Y_1-Y_2)}}{\ln\frac{1}{r^2\Lambda^2}}\bigg)+ \frac{B'\Nc}{b\pi\ln\frac{1}{r^2\Lambda^2}}-\frac{1}{2}\bigg]\nn
&&\hspace*{1.0cm}\times c(Y-Y_0-Y_1-Y_2)+\frac{2B'\Nc^2}{b^2c\pi^2}\sqrt{c(Y-Y_0-Y_1-Y_2)}\Bigg\},
\label{GF_R}
\eea
and
\bea
\mathbb{S}_L(r_1,Y_2) &=& \exp\Bigg\{-\frac{\Nc}{bc\pi}\bigg[\ln\bigg(\frac{\sqrt{c(Y_1+Y_2)}}{\ln\frac{1}{r^2\Lambda^2}}\bigg)+ \frac{B'\Nc}{b\pi\ln\frac{1}{r^2\Lambda^2}}-\frac{1}{2}\bigg]c(Y_1+Y_2)
+\frac{2B'\Nc^2}{b^2c\pi^2}\sqrt{c(Y_1+Y_2)}\nn
&&\hspace*{1.0cm}+\frac{\Nc}{bc\pi}\bigg[\ln\bigg(\frac{\sqrt{cY_1}}{\ln\frac{1}{r^2\Lambda^2}}\bigg)+ \frac{B'\Nc}{b\pi\ln\frac{1}{r^2\Lambda^2}}-\frac{1}{2}\bigg]cY_1
-\frac{2B'\Nc^2}{b^2c\pi^2}\sqrt{cY_1}\Bigg\}.
\label{GF_L}
\eea
By substituting Eqs.(\ref{Sdipole_gf})-(\ref{GF_L}) into Eq.(\ref{Srareloge}), we get
\bea
S(r_0,r_1,Y) &=& \exp\Bigg\{-\frac{\Nc}{bc\pi}\bigg[\ln\bigg(\frac{\sqrt{c(Y-Y_0-Y_1-Y_2)}}{\ln\frac{1}{r^2\Lambda^2}}\bigg)+ \frac{B'\Nc}{b\pi\ln\frac{1}{r^2\Lambda^2}}-\frac{1}{2}\bigg]c(Y-Y_0-Y_1-Y_2)\nn
&&\hspace*{1.0cm}+\frac{2B'\Nc^2}{b^2c\pi^2}\sqrt{c(Y-Y_0-Y_1-Y_2)}
-\frac{\Nc}{bc\pi}\bigg[\ln\bigg(\frac{\sqrt{c(Y_1+Y_2)}}{\ln\frac{1}{r^2\Lambda^2}}\bigg)+ \frac{B'\Nc}{b\pi\ln\frac{1}{r^2\Lambda^2}}-\frac{1}{2}\bigg]c(Y_1+Y_2)\nn
&&\hspace*{1.0cm}+\frac{2B'\Nc^2}{b^2c\pi^2}\sqrt{c(Y_1+Y_2)}\Bigg\}S(r, Y_0),
\label{GF_comb}
\eea
where the unknown variable $Y_1$ can be determined by maximizing the $S$-matrix, or equivalently via minimizing of the exponent of Eq.(\ref{GF_comb}). One gets
\be
Y_1 = \frac{1}{2}(Y-Y_0) - Y_2.
\label{Y1}
\ee
We substitute $Y_1$ into Eq.(\ref{GF_comb}) and obtain
\bea
S(r,Y) &=& \exp\Bigg\{-\frac{\Nc}{bc\pi}\bigg[\ln\bigg(\frac{\frac{\sqrt{2}}{2}\ln\frac{Q_s^2}{\Lambda^2}}{\ln\frac{1}{r^2\Lambda^2}}\bigg)+ \frac{B'\Nc}{b\pi\ln\frac{1}{r^2\Lambda^2}}-\frac{1}{2}\bigg]\ln^2\frac{Q_s^2}{\Lambda^2}
+\frac{2\sqrt{2}B'\Nc^2}{b^2c\pi^2}\ln{\frac{Q_s^2}{\Lambda^2}}\Bigg\}\nn
&&\hspace*{1.0cm}\times S(r, Y_0),
\eea
which is exactly the same as the Eq.(\ref{SNLLrareres}) obtained in the center of mass frame.

%-----------------------------------------------------------------------------

\end{document}